\documentclass[aps,preprint,amsmath,amssymb]{revtex4}
\usepackage{graphicx,color}
\def\be{\begin{eqnarray}}
\def\ed{\end{eqnarray}}
\def\non{\nonumber}
\def\ga{\gamma}

\def\U{\cal U}
\def\l{\lambda}
\def\hs{\hat s}
\def\hatt{\hat \theta}
\def\la{\langle}
\def\ra{\rangle}

\begin{document}

\preprint{USM-TH-274}

\title{\Large \bf   Forward-backward asymmetry of top quark \\
in unparticle physics }

\date{\today}

\author{\bf
Chuan-Hung~Chen$^{1,2}$\footnote{E-mail: physchen@mail.ncku.edu.tw
}, G.~Cveti\v{c}$^{3}$\footnote{E-mail: gorazd.cvetic@usm.cl} and C.~S.~Kim$^{4}$\footnote{E-mail: cskim@yonsei.ac.kr,
~~ Corresponding Author} }
\affiliation{
$^{1}$ Department of Physics, National Cheng-Kung University, Tainan 701, Taiwan \\
$^{2}$National Center for Theoretical Sciences, Hsinchu 300, Taiwan
\\
$^{3}$Department of Physics and Valparaiso Center for Science and Technology, Universidad T\'ecnica Federica Santa Mar\'{\i}a, Valpara\'{\i}so, Chile
\\
$^{4}$ Department of Physics $\&$ IPAP, Yonsei University, Seoul 120-479, Korea
}

\begin{abstract}
\noindent The updated CDF measurement of the forward-backward asymmetry $A_{{\rm FB}}$ in the top quark
production $p{\bar p} \to t{\bar t}$ at Tevatron (with $\sqrt{s}=$ 1.96 TeV) shows a deviation of
$2 \sigma$ from the value predicted by the Standard QCD Model. We present calculation of this quantity
in the scenario where colored unparticle physics contributes to the s-channel of the process,
and obtain the regions in the plane of the unparticle parameters $\lambda$ and $d_{\U}$,
which give the values of the $A_{{\rm FB}}$ and of the total $t{\bar t}$ production cross section compatible
with the present measurements.
\end{abstract}

\maketitle

Due to C-parity invariance, it is known that the forward-backward asymmetry (FBA)
of top quark pair production at Tevatron vanishes at leading order (LO) \cite{Halzen:1987xd}
in the standard QCD model (SM). The inclusive non-zero charge asymmetry
can be induced by (i) radiative corrections to
quark-antiquark annihilation and (ii) interference between different amplitudes contributing to
gluon-quark scattering $q g \to t \bar t q$ and $\bar q g \to t \bar t \bar q$ \cite{Halzen:1987xd,Kuhn:1998jr}.
Within the SM, this leads, at the Tevatron with $\sqrt{s}=1.96$ TeV, to the nonzero but relatively low prediction \cite{AKR},
\begin{equation}
A^{p\bar p}_{\rm FB}({\rm SM}) =0.050 \pm 0.015 \ .
\label{FBSM}
\end{equation}
Measurement of any significant deviation from this SM prediction could be
attributed to the new physics effects.

When D0 Collaboration published the first measurement on the FBA in
top-quark pair production in the $p\bar p$ laboratory frame with
0.9 fb$^{-1}$ of data, an unexpectedly larger FBA value was
indicated \cite{D0_PRL100}. By using 1.9 fb$^{-1}$ \cite{CDF_PRL101},
the CDF Collaboration observed the asymmetry (at $\sqrt{s}=1.96$ TeV) to be
 \be
 A^{p\bar p}_{\rm FB}=0.17 \pm 0.08 \ , \qquad A^{t\bar t}_{\rm FB} =
0.24 \pm 0.14\,,
 \ed
in the $p\bar p$ frame and $t\bar t$ frame, respectively. The updated CDF
result with luminosity of $3.2$ fb$^{-1}$, in the $p\bar p$ (lab) frame,
is \cite{CDFnote}
 \be
 A^{p\bar p}_{{\rm FB}}({\rm exp})=
0.193 \pm 0.065(\rm stat) \pm 0.024 (\rm syst) \ .
\label{AFB}
 \ed
As seen in Eq.~(\ref{AFB}), the large value of the FBA of top-quark is not
smeared by the statistics. Inspired by the $2 \sigma$ deviation of the
observed value of the top quark FBA from the SM-predicted value, several
possible solutions have been proposed and studied by authors in
Refs.~\cite{Djouadi:2009nb,Ferrario:2009bz,Jung:2009jz,Cheung:2009ch,Frampton:2009rk,Shu:2009xf,Arhrib:2009hu,Ferrario:2009ee,Dorsner:2009mq,Jung:2009pi,Cao:2009uz,Barger:2010mw,Cao:2010zb,Kumar:2010vx,Martynov:2010ed,Chivukula:2010fk,Bauer:2010iq}.

In general, the top-pair production by the new physics could be through
s-, t- and u-channel and the situation depends on the property of the new
particle. No matter to which channel they contribute, the extensions of the
SM in the framework of particle physics, such as axigluon
\cite{Ferrario:2009bz,Frampton:2009rk}, $Z'$ \cite{Jung:2009jz},
$W'$ \cite{Cheung:2009ch}, diquarks \cite{Arhrib:2009hu}, etc., have some
drawbacks. For instance,  in order to explain the observed top-quark
FBA value, one has to introduce unimaginably large flavor-changing couplings
in the t- and u-channels. The couplings in the s-channel could be as large
as the strong gauge coupling of the SM. However, beside the serious
constraint from the invisible production of a new resonance, the sign of the
top-quark couplings has to be chosen opposite to that of the light quarks in
order to get the correct sign of the FBA value. In order to avoid the
aforementioned problems, we study in this work the top quark FBA in the
framework of unparticle physics which is dictated by the scale of conformal
invariance.

An exact scale-invariant ``stuff'' cannot have a definite mass unless it is
zero. Therefore, in order to distinguish it from the conventional particles,
Georgi named the ``stuff'' unparticle \cite{Georgi1, Georgi2}.
It was found that the unparticle has a noninteger scaling
dimension $d_{\U}$ and behaves as an invisible particle \cite{Georgi1}
(see also \cite{Gaete:2008wg}).
Further implications of the unparticle to collider and low energy physics are
discussed in Refs.~\cite{un1,un2,un3}. We will adopt three aspects of
unparticle physics in order to present a possible explanation of the
aforementioned large value of the FBA. Firstly, if we take the protecting
symmetry to be exact, then, due to the unique character of indefinite mass,
no visible resonant unparticle will be produced in the
$p\bar p$ collisions. Secondly, by utilizing the noninteger scale dimension,
the differential cross section for $t\bar t$ production could be enhanced
without fine-tuning the large couplings of unparticle and quarks. Finally, to
match the interaction structure of the SM, the considered scale invariant
stuff (unparticle) is a vector boson and carries color charges
\cite{Cacciapaglia:2007jq}; it has chiral couplings to
quarks and its representation in $SU(3)_c$ belongs to color-octet.

Since there is no well established approach to give a full theory for
unparticle interactions, we study instead the topic from the phenomenological
viewpoint. In order to escape the large couplings from flavor changing
neutral currents (FCNCs), the couplings of unparticle to quarks are chosen to
be flavor conserving. Hence, we write the interactions of colored unparticle
with quarks as
 \be
 {\cal L} &=& \bar q (g^q_V \ga_\mu + g^q_A \ga_\mu \ga_5) T^a q {\cal O}^{a\mu}_{\U} \ ,
\label{eq:lag_un}
 \ed
where $g_{\chi}=\l^q_\chi/\Lambda^{d_{\U}-1}_{\U}$ and $\chi= V$ and $A$. Here,
$\l^q_\chi$ is the dimensionless coupling and the index $q$ denotes the quark
flavor, $\Lambda_{\U}$ is the scale at
which the unparticle 
is formed,
and $\{T^a\} = \lambda^a/2$ are the $SU(3)_c$ generators (where $\lambda^a$
are the Gell-Mann matrices) normalized by $tr (T^a T^b) = \delta^{ab}/2$. The
power $d_{\U}-1$ is determined from the effective Lagrangian of
Eq.~(\ref{eq:lag_un}) in four-dimensional spacetime when the dimension of
the colored unparticle ${\cal O}^{q\mu}_{\U}$ is taken as $d_{\U}$.
By following the scheme shown in Ref.~\cite{Grinstein:2008qk},
the propagator of the colored vector unparticle is written as
 \be
&& \int d^4x e^{-ik\cdot x} \la 0|T {\cal O} ^{a}_{\mu}(x) {\cal O}^{b}_{\nu}(0)|0\ra  \non \\
&=& -iC_V \frac{ \delta^{ab}}{(-p^2 -i\epsilon)^{3-d_{\U}}} \left[p^2 g_{\mu \nu} -\frac{2(d_{\U}-2)}{d_{\U}-1}p_\mu p_\nu \right] \label{eq:prop_un}
 \ed
with
 \be
 C_V &=& \frac{A_{d_{\U}}}{2\sin d_{\U} \pi} \,, \non\\
 A_{d_{\U}}&=& \frac{16\pi^{5/2}}{(2\pi)^{2d_{\U}}} \frac{\Gamma(d_{\U} +1/2)}{\Gamma(d_{\U}-1) \Gamma(2d_{\U})}\,.
 \ed

After introducing the interactions of unparticle with quarks and the virtual
unparticle propagator, we can now calculate the $t\bar t$ pair production at
the quark level. Using Eqs.~(\ref{eq:lag_un}) and (\ref{eq:prop_un}), the
scattering amplitude for $q \bar q\to t\bar t$ by unparticle exchange in the
s-channel is
 \be
A_{\U} &=&  \bar q \left( g^q_V \ga_\mu + g^q_A \ga_\mu \ga_5 \right) T^a q\frac{C_V}{(-p^2-i\epsilon)^{3-d_{\U}}} \left[ p^2 g_{\mu\nu} -\frac{2(d_{\U}-2)}{d_{\U}-1}p_\mu p_\nu \right]\non \\
&\times&
  \bar t \left( g^t_V \ga_\nu + g^t_A \ga_\nu \ga_5 \right) T^a t
 \ed
where flavor $q$ denotes the light $u$ and $d$ quark, and
$p=p_q + p_{\bar q} = p_t + p_{\bar t}$. The t-channel does not contribute, due
to flavor-conserving vertices Eq.~(\ref{eq:lag_un}). The equations of motion imply 
${\bar q} {\slash \!\!\!p} q =0$ and $\bar q {\slash  \!\!\!p} \ga_5 q =-2m_q \bar q \ga_5  q$. Thus, the factor $2(d_{\U}-2)/(d_{\U}-1)$ in the propagator is associated with the light quark mass and is negligible. Consequently, the scattering amplitude combined with the SM contributions is given by
 \be
 A &=& A_{{\rm SM}} + A_{\U} \non\\
 &=& \frac{g^2_s}{\hs}\bar q \ga_\mu T^a q \bar t \ga^\mu T^a t \non\\
 &+& \frac{ \hs C_V }{\hs^{3-d_{\U}}}e^{-i\pi(3-d_{\U})} \bar q \left( g^q_V \ga_\mu + g^q_A \ga_\mu \ga_5 \right) T^a q \bar t \left( g^t_V \ga^\mu + g^t_A \ga^\mu \ga_5 \right) T^a t \ ,
\label{ampl}
 \ed
with $g_s$ being the strong coupling of the QCD SM and
$\hs=(p_q + p_{\bar q})^2 = (p_t + p_{\bar t})^2$.
For explicitly showing the differential cross section in $t\bar t$ invariant
mass frame, we choose  the relevant coordinates of particle momenta as
 \be
 &&p_{q,\bar{q}}= \frac{\sqrt{\hs} }{2}(1,0,0,\pm 1)\,, \non\\
&&p_{t,\bar{t}}= \frac{\sqrt{\hs} }{2}(1,\pm\beta_t \sin\hatt,0,\pm
\beta_t \cos\hatt)\,,
  \ed
with $\beta^2_t = 1-4m^2_t/\hat{s}$. The polar angle $\hat\theta$ is the relative
angle between outgoing top-quark and the incoming $q$-quark.  The spin and
color averaged amplitude-square is straightforwardly obtained as
 \be
 |\bar A|^2 &=& \frac{1}{2^2} \frac{1}{N^2_C} |A|^2\,, \non \\
  &=&  \frac{(N^2_C -1)}{16N^2_C} \left\{ 4 (4\pi \alpha_s)^2 \left( 1+\beta^2_t \cos^2\hatt + 4 \frac{m^2_t}{\hs} \right) \right.\non \\
&+& 8(C_V 4\pi \alpha_s) \cos\pi(3-d_{\U}) \frac{\hs^2}{\hs^{3-d_{\U}}} \left[g^q_V g^t_V \left( 1+\beta^2_t \cos^2\hatt + 4 \frac{m^2_t}{\hs}\right) + 2g^q_A g^t_A \beta_t \cos\hatt \right]  \non \\
&+& 4\hs^2\left( \frac{\hs C_V}{\hs^{3-d_{\U}}}\right)^2 \left[
 (g^{t}_{V})^2 \left( (g^{q}_V)^2 + (g^{q}_{A})^2 \right) \left(1+\beta^2_t \cos^2\hatt + \frac{4m^2_t}{\hs} \right)\right.\non \\
&+& \left.\left. (g^{t}_{A})^2 \left( (g^{q}_V)^2 + (g^{q}_{A})^2 \right) \left(1+\beta^2_t \cos^2\hatt - \frac{4m^2_t}{\hs} \right)+ 8 g^q_V g^t_V g^q_A g^t_A \beta_t \cos\hatt \right]
\right\}\,.
 \ed
As a consequence,  the differential cross section for $q\bar q\to t\bar t$ process as a function of $\hat\theta$ in $t\bar t$ frame is found to be
 \be
 \frac{d\hat{\sigma}^{q\bar q \to t\bar t}}{d\cos\hatt} &=& \frac{N^2_C -1}{128 N^2_C \pi \hs} \beta_t\left\{ (4\pi \alpha_s)^2 \left( 1+\beta^2_t \cos^2\hatt + 4 \frac{m^2_t}{\hs} \right) \right.\non \\
 &+& 2C_V (4\pi\alpha_s) \cos\pi(3-d_{\U}) \frac{\hs^2}{\hs^{3-d_{\U}}}\left[g^q_V g^t_V \left( 1+\beta^2_t \cos^2\hatt + 4 \frac{m^2_t}{\hs}\right) + 2g^q_A g^t_A \beta_t \cos\hatt \right] \non \\
 &+& \left(\frac{\hs^2 C_V}{\hs^{3-d_{\U}}}\right)^2 \left[
 (g^{t}_{V})^2 \left( (g^{q}_V)^2 + (g^{q}_{A})^2 \right) \left(1+\beta^2_t \cos^2\hatt + \frac{4m^2_t}{\hs} \right) \right.\non \\
&+& \left.\left. (g^{t}_{A})^2 \left( (g^{q}_V)^2 + (g^{q}_{A})^2 \right) \left(1+\beta^2_t \cos^2\hatt - \frac{4m^2_t}{\hs} \right)+ 8 g^q_V g^t_V g^q_A g^t_A \beta_t \cos\hatt \right]
\right\}\,.
\label{eq:angle}
 \ed
In the s-channel, only the terms linear in $\cos\hat\theta$ will contribute 
directly to the forward-backward asymmetry.
From Eq.~(\ref{eq:angle}), the relevant effects are associated with
$g^q_{A} g^t_{A}$ and $g^{q}_{V} g^t_V g^q_A g^t_A$, in which the former is from
the interference between unparticle and SM while the latter is from the
contribution of unparticle itself. In both terms, we see clearly that the
axial-vector couplings are the essential to generate the FBA.

To obtain the hadronic cross section from the parton level,  we have to
consider the convolution  with the parton distribution functions. Thus, the
differential cross section at the hadronic level is
 \be
 \frac{d\sigma (p{\bar p} \to t{\bar t})}{d\cos\theta} &=& \sum_{ij}
\int_{x_2=0}^1 \int_{x_1=0}^1 dx_1 d x_2 f_i(x_1) f_j(x_2) \frac{\partial \hat{\sigma}^{q_i \bar q_j\to t\bar t}(\theta,x_1,x_2) }{\partial \cos \theta} \ ,
\label{eq:dcross}
 \ed
where $f_i$ ($f_j$) is the parton distribution function of the parton $q_i$
(${\bar q}_j$) in the proton (antiproton), the angle $\theta$ represents the
angle between the three-momentum of the produced $t$ quark and the
three-momentum of the proton $p$ ($\Leftrightarrow$ of the quark $q$)
in the lab system (center of mass system of $p{\bar p}$). The sum over $(i,j)$
in Eq.~(\ref{eq:dcross}) is over all parton pair combinations
$q {\bar q} = q_i {\bar q}_j$ for the scattering process $q_i {\bar q}_j \to t {\bar t}$
($q_i, q_j = u,d,s$).

In the following, all the unprimed kinematic quantities are in the lab
system, and all the ``hatted'' kinematic quantities are in the center of
mass system (CMS) of $q {\bar q}$ ($ \Leftrightarrow$ CMS of $t {\bar t}$). Taking
into account the relations $p_q=x_1 p_p$ and $p_{\bar q} = x_2 p_{\bar p}$ in the
lab system, considering the four-momentum conservation in the scattering
$q {\bar q} \to t {\bar t}$ in the $q {\bar q}$ CMS, and relating the lab and
$q {\bar q}$ CMS quantities via the corresponding boost relations, the
following relation can be obtained between the angle $\theta$ and its
$q {\bar q}$ CMS analog ${\hat \theta} ={\hat \theta}(\theta,x_1,x_2)$:
\begin{eqnarray}
\cos \left( \hat \theta (\theta,x_1,x_2) \right) & = & \frac{1}{\beta_t
\left[ (x_1+x_2)^2 - \cos^2 \theta (x_1 - x_2)^2 \right] } {\bigg \{} - (x_1^2 - x_2^2) \sin^2 \theta
\nonumber\\
&&
+ 4 \cos \theta \left[ x_1^2 x_2^2 \beta_t^2 - \frac{m_t^2}{s} (x_1 - x_2)^2 \sin^2 \theta \right]^{1/2} {\bigg \}} \ ,
\label{cthatth}
\end{eqnarray}
where $\beta_t$ is the aforementioned quantity involving ${\hat s} = (p_q + p_{\bar q})^2 = x_1 x_2 s$
\begin{equation}
\beta_t = \beta_t(x_1 x_2) = \sqrt{ 1 - \frac{m_t^2}{x_1 x_2 s} } \ .
\label{bt}
\end{equation}
The relevant independent kinematic quantities in the integration (\ref{eq:dcross}) are all lab-related: $x_1$, $x_2$, and $\theta$. On the other hand, Eq.~(\ref{cthatth}) shows that the $q {\bar q}$ CMS-related angle $\hat \theta$ is a function of the aforementioned three independent quantities $x_1$, $x_2$, and $\theta$. The quantity $\partial {\hat{\sigma}}^{q \bar q \to t\bar t}/\partial \cos \theta$ appearing as integrand in Eq.~(\ref{eq:dcross}) is obtained directly from Eqs.~(\ref{eq:angle}) and (\ref{cthatth}) by applying the derivatives at fixed $x_1$ and $x_2$
\begin{equation}
 \frac{\partial \hat{\sigma}^{q \bar q\to t\bar t} (\theta, x_1, x_2)}{\partial \cos \theta} =
 \frac{d\hat{\sigma}^{q \bar q\to t\bar t}}{d\cos \hat \theta} \
\frac{ \partial \cos \left( \hat \theta(\theta,x_1,x_2) \right)}{\partial \cos \theta} \ ,
\label{dhatsigdcth}
\end{equation}
where the partial derivatives $\partial/\partial \cos \theta$ are at fixed $x_1$ and $x_2$.

The total hadronic cross section $\sigma(p{\bar p} \to t{\bar t})$
and the corresponding forward-backward asymmetry are then obtained by the
corresponding integrations of the expression (\ref{eq:dcross}) in the lab frame
\begin{eqnarray}
\sigma(p{\bar p} \to t{\bar t}) & = &
\int_{-1}^1 d \cos \theta \; \frac{d\sigma (p{\bar p} \to t{\bar t})}{d\cos\theta} \ ,
\label{sigmatot}
\\
 A^{p\bar p}_{{\rm FB}} &=& \left(
\int_0^{1}  d\cos\theta \; \frac{d\sigma (p{\bar p} \to t{\bar t})}{d\cos\theta}
- \int_{-1}^{0}  d\cos\theta \; \frac{d\sigma (p{\bar p} \to t{\bar t})}{d\cos\theta} \right)/\sigma(p{\bar p} \to t{\bar t}) \ .
\label{Afb} \
\end{eqnarray}
Another physical observable of experimental interest is the invariant
mass distribution $d \sigma/d M_{t {\bar t}}$ \cite{Aaltonen:2009iz} where
$M^2_{t \bar t} = (p_t + p_{\bar t})^2 = x_1 x_2 s$
\begin{equation}
\frac{d \sigma(p{\bar p} \to t {\bar t})}{d M_{t {\bar t}}} =
2 \ \frac{M_{t {\bar t}}}{s} \int_{M^2_{t {\bar t}}/s}^1  \frac{d x_1}{x_1} \sum_{i,j}
f_i(x_1) f_j(x_2) \int^{1}_{-1} d\cos\theta \frac{\partial
\hat{\sigma}^{q_i \bar q_j\to t\bar t}(\theta,x_1,x_2)}
{\partial \cos \theta}{\bigg |}_{x_2=M^2_{t {\bar t}}/(s x_1)} \ .
\label{dsigdMtt}
\end{equation}
The integrations in Eqs.~(\ref{sigmatot})-(\ref{dsigdMtt})
are performed across the kinematically allowed regions, $i.e.$, such that
$\beta_t$ and $\cos \hat \theta$ are real.

After having presented formulas for the three physical observables, we can
now numerically investigate the unparticle contributions to top-quark pair
production. At first, in order to reduce the number of free parameters, we
assume that the colored vector unparticle is flavor  blind, $i.e.$
$g^t_V=g^t_A =g^q_V =g^q_A = g$. Then,
the remaining unknown parameters appearing
in the physical quantities are:
$g = \lambda/\Lambda_{\U}^{d_{\U}-1}$, and the scale dimension $d_{\U}$.
This means that
in such a case we have only two independent parameters $\lambda$ and $d_{\U}$,
both dimensionless quantities, and we can fix the scale $\Lambda_{\U}$
formally to an arbitrary value. We will set it equal to $\Lambda_{\U}=1$ TeV.
We will see that, unlike the situation in the axigluon model \cite{Ferrario:2009bz,Frampton:2009rk}
in which $g^q_A = -g^t_A$ is necessary to get the positive sign in FBA,
in unparticle physics {\it the flavor-blind and chirality-independent couplings} are enough to fit the data.

Further, it is necessary to consider the measurements of at least two
of the aforementioned three observables, namely $\sigma(p{\bar p} \to t{\bar t})$
and $A_{{\rm FB}}^{p{\bar p}}$, Eqs.~(\ref{sigmatot})-(\ref{Afb}), in order to
restrict the area of the parameters $\lambda$ and $d_{\U}$. The value of the
$t {\bar t}$ production cross section $\sigma(p{\bar p} \to t{\bar t})$ was measured by
the CDF Collaboration \cite{Abazov:2009ae}\
\begin{eqnarray}
 \sigma (p\bar p \to t \bar t)^{\rm exp} &=& 7.50\pm 0.31\;({\rm stat})\; \pm 0.34\;({\rm
   syst}) \;\pm 0.15 \; ({\rm th}) \;  \ \ {\rm pb}
\nonumber\\
& = & 7.50 \pm 0.48 \; {\rm pb} \ .
\label{sigtotCDF}
\end{eqnarray}
On the other hand, the SM prediction is
$\sigma(p\bar p\to t\bar t)^{\rm SM}=6.73^{+0.71}_{-0.79}$ pb \cite{ttnlo},
which includes the contributions from the tree-level,
the next-to-leading order in $\alpha_s$,
and the next-to-leading in threshold logarithms (LO+NLO+NLL).
In the specific case of using the CTEQ6.6 parton distribution functions
\cite{CTEQ66} (which we use), the central value for the SM prediction
goes slightly down to
$\sigma(p\bar p\to t\bar t)^{\rm SM} = 6.61$ pb, Ref.~\cite{ttnlo}
(Cacciari {\it et al.\/}, 2008) when the top quark (pole) mass is taken
to be $m_t=175$ GeV.

The other measurement is the aforementioned FBA value Eq.~(\ref{AFB}). The
NLO effects in the QCD SM give nonzero FBA value $0.050 \pm 0.015$,
Eq.~(\ref{FBSM}).
However, in our calculations, the SM amplitude is the tree-level amplitude
[rescaled accordingly in order to obtain
$\sigma(p{\bar p} \to t{\bar t})^{\rm SM}=6.6$ pb, see below], which gives
$A_{\rm FB}^{p{\bar p}}=0$. Therefore, we will regard the $A_{\rm FB}^{p{\bar p}}$
as calculated according to Eq.~(\ref{Afb}) [using Eqs.~(\ref{dhatsigdcth}),
(\ref{eq:dcross}) and (\ref{eq:angle})] to be responsible for the deviation
of the experimental from the SM FBA value
 \be
 A^t_{{\rm FB}} \equiv A^{p\bar p}_{{\rm FB}}({\rm exp})-
A^{p\bar p}_{{\rm FB}}({\rm SM}) = 0.143 \pm 0.071 \ ,
\label{Afbour}
 \ed
where the uncertainties ($\pm 0.065$, $\pm 0.024$, $\pm 0.015$) were added in
quadrature.

\begin{figure}[t]
\includegraphics*[width=4.5 in]{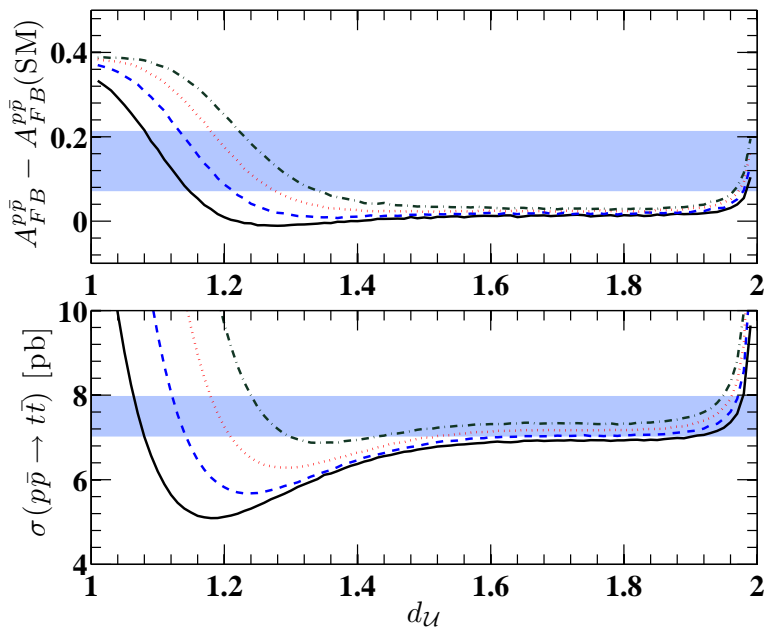}
\caption{ $t\bar t$ production cross section (the lower) and top-quark
FBA (the upper figure) as a function of the scale dimension $d_{\cal U}$,
where the solid, dashed, dotted and
dash-dotted lines represents $\lambda$=1.4, 1.6, 1.8, 2.0, respectively.
The band in the plot represents the measured values
with 1$\sigma$ uncertainties. }
 \label{fig:lam}
\end{figure}

\begin{figure}[t]
\includegraphics*[width=4.5 in]{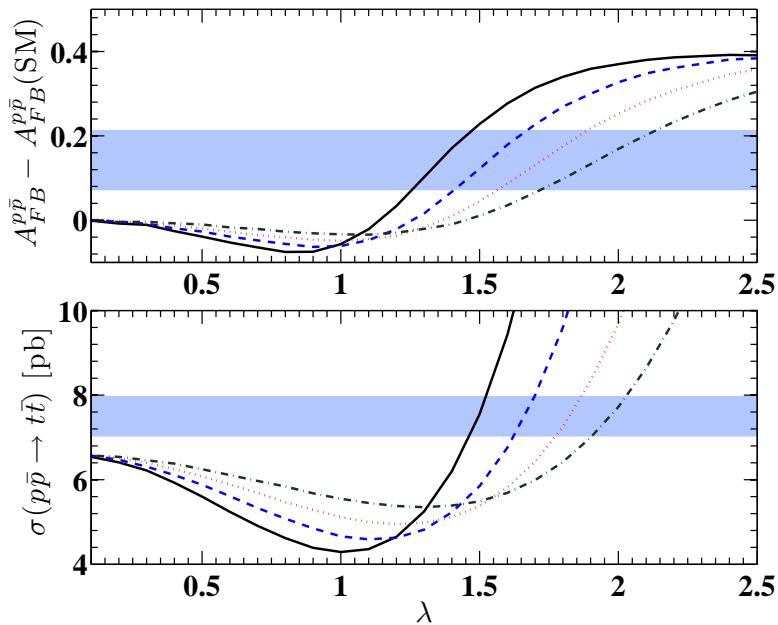}
\caption{ As Fig.~\ref{fig:lam}, but as a function of the parameter
$\lambda$. The solid, dashed, dotted and dash-dotted lines represents
$d_{\cal U}$=1.1, 1.15, 1.2, 1.25, respectively. }
 \label{fig:du}
\end{figure}

In our calculations we use the CTEQ6.6 parton distribution functions,
the value $m_t=175$ GeV for the $t$ quark (pole) mass, and for the QCD
coupling the value $\alpha_s \approx \alpha_s(m_t) \approx 0.11$. With such values,
we obtain the tree-level $\sigma({\rm SM;tree}) \approx 4.85$ pb. We use for the SM
amplitude the rescaling factor $\sqrt{1.36}$ in order to obtain
$\sigma({\rm SM}) = 6.6$ pb, which is according to Ref.~\cite{ttnlo}
(the second entry: Cacciari {\it et al.\/}, 2008) the central value
of $\sigma({\rm SM})$ when $m_t=175$ GeV and CTEQ v6.6 is used for the parton
distribution functions.

\begin{figure}[t]
\includegraphics*[width=4.5 in]{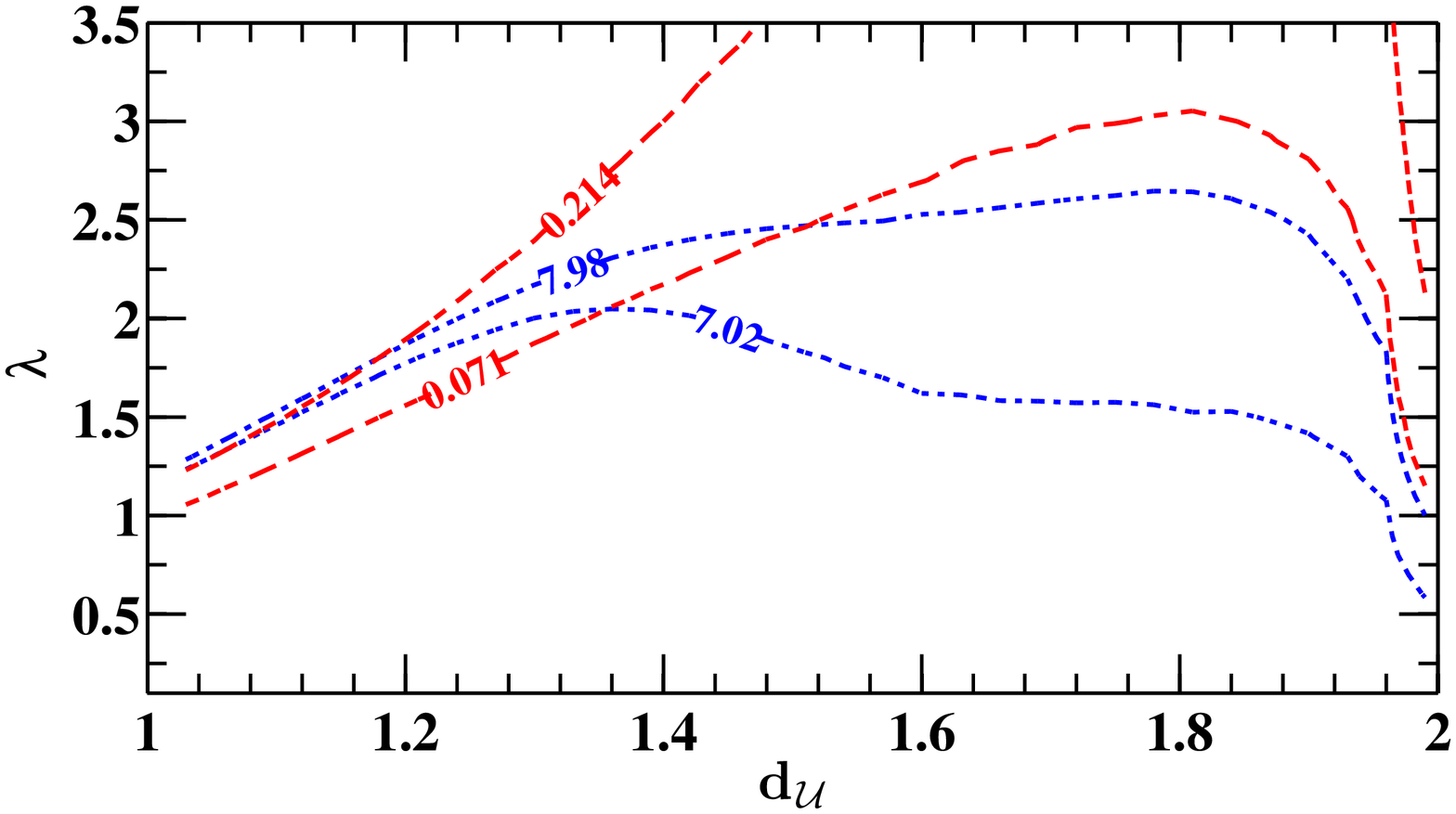}
\caption{  Contours for $\sigma(p\bar p\to t\bar t)$ (dotted) and
$A^{p\bar p}_{{\rm FB}}-A^{p\bar p}_{{\rm FB}}({\rm SM})$ (dashed) as a function of
$d_{\cal U}$ and $\lambda$. The numbers in the plot denote the lower and upper
bounds of each observable with $1\sigma$ uncertainties of the data:
$\sigma = 7.50 \pm 0.48$ pb; $A_{\rm FB}^{p{\bar p}}$ as in Eq.~(\ref{Afbour}). }
 \label{fig:lam_du}
\end{figure}

\begin{figure}[t]
\includegraphics*[width=4.5 in]{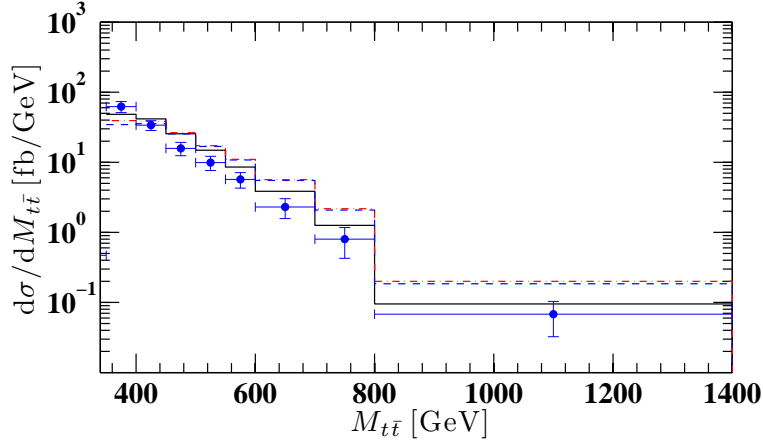}
\caption{ $d\sigma/dM_{t\bar t}$ as a function of invariant mass of top-pair
$M_{t\bar t}$, where the solid, dash-dotted and dashed lines represent the
SM result and colored unparticle with $(\lambda,\, d_{\cal U})=(2.05, 1.28)$
and $(1.70,\, 1.175)$, respectively.   The vertical bars are the data from CDF
measurement with an integrated luminosity of 2.5 fb$^{-1}$,
Ref.~\cite{Aaltonen:2009iz}. }
 \label{fig:Mtt}
\end{figure}

\begin{figure}[t]
\includegraphics*[width=4.5 in]{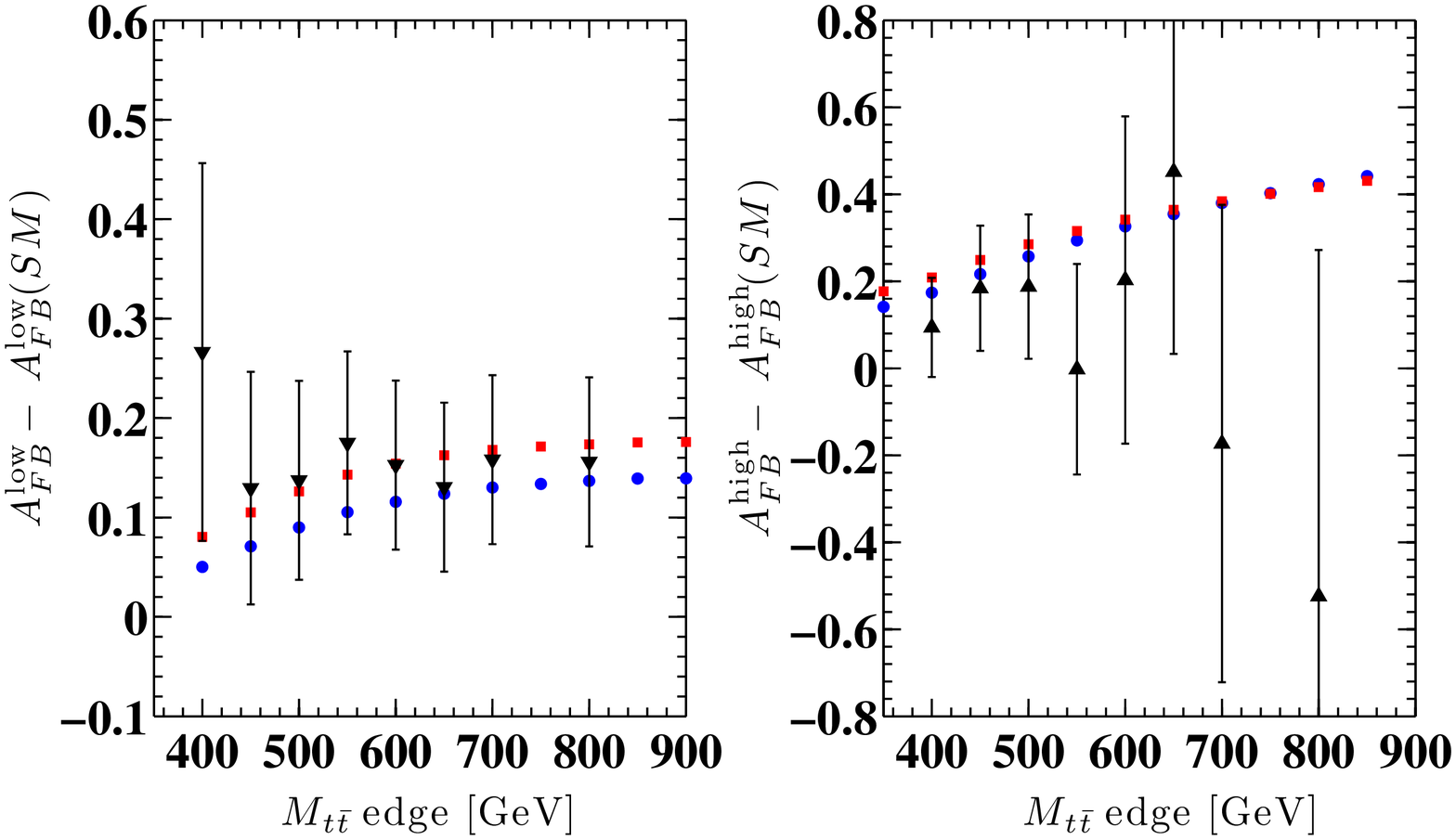}
\caption{ Restricted forward-backward asymetries
$A_{\rm FB}^{t,{\rm low}}$ and $A_{\rm FB}^{t,{\rm high}}$ as functions of the
threshold (``edge'') $M_{t {\bar t}}$ values, for
$(\lambda,\, d_{\cal U})=(2.05, 1.28)$ (circles)
and $(1.70,\, 1.175)$ (squares).
Included are also the corresponding CDF measured values \cite{CDF9724} (their 8th and 9th figure)
subtracted by the SM values \cite{Almeida:2008ug}, as bars with triangles.}
\label{fig:AFBLH}
\end{figure}

The physically interesting regime for unparticle physics is $1 < d_{\U} < 2$, and
$\lambda \stackrel{<}{\sim} 10^0$. We calculated $A_{{\rm FB}}$,
$\sigma$ and $d\sigma/dM_{t {\bar t}}$, scanning over the parameter regions
$0 < \lambda < 3.5$ and $1 < d_{\U} < 2$. The numerical results are presented in
Figs.~\ref{fig:lam}-\ref{fig:AFBLH}.
We show the $t\bar t$ production cross section and $\left[ A^{p\bar p}_{FB}-A^{p\bar p}_{FB}({\rm SM}) \right]$
as a function of $d_{\U}$ ($\lambda$) in Figs.~\ref{fig:lam} (\ref{fig:du}),
where the solid, dashed, dotted and dash-dotted lines represents $\lambda=1.4, 1.6, 1.8, 2.0$
($d_{\U}=1.1, 1.15, 1.2, 1.25$), respectively.
Figure \ref{fig:lam_du} is the central
result of our calculations. It shows the region in the $d_{\U}$-$\lambda$
parameter plane  which simultaneously fulfills the experimental constraints
(\ref{sigtotCDF}) and (\ref{Afbour}). This region lies between the two
dotted and simultaneously between the two dashed lines. The central measured
values $\sigma \approx 7.5$ pb and $A^{p\bar p}_{{\rm FB}}({\rm exp})-
A^{p\bar p}_{{\rm FB}}({\rm SM}) \approx 0.14$ are achieved at $\lambda=2.05$ and
$d_{\U}=1.28$.
In Fig.~\ref{fig:lam_du}, we scanned over the
free parameter space in finite steps $\Delta \lambda = 0.1$ and
$\Delta d_{\U}=0.01$.

In Fig.~\ref{fig:Mtt} we present the average values of $d \sigma/dM_{t {\bar t}}$
in eight different $M_{t {\bar t}}$-intervals (``bins'') as used by the CDF
measurement \cite{Aaltonen:2009iz}. The presented results are for: (a) the
QCD SM case (solid line: $\lambda=0$; $\sigma(t {\bar t}) = 7.5$ pb);
(b) the ``central'' case (dash-dotted line; $\lambda=2.05$ and $d_{\U}=1.28$,
giving $\sigma(t {\bar t})=7.5$ pb and FBA of Eq.~(\ref{Afbour}) equal 0.14);
(c) another, more ``marginal'' case
(dashed line; $\lambda=1.70$ and $d_{\U}=1.175$,
giving $\sigma(t {\bar t}) = 7.01$ pb and FBA of Eq.~(\ref{Afbour}) equal 0.178).
The CDF measurements are the points
with vertical lines. All our results (including the QCD SM) are above the
CDF measurements, with the exception of the first two bins
$M_{t {\bar t}} \leq 450$ GeV.
The deviations could plausibly be ascribed to two main uncertainties \cite{Frederix:2007gi}:

\begin{itemize}
\item (i) The chosen scales of renormalization ($\mu_R$) and factorization ($\mu_F$)
for which the usual possible values could be taken between $m_t/2$ and $2 m_t$. Here we adopted $\mu_R=\mu_F=m_t$.
\item (ii) The $M_{t\bar t}$-dependent NLO effects which include the NLO parton distribution fuction (PDF).
Here for simplicity we just use a $M_{t\bar t}$-independent scale factor value of K=1.36
($i.e.$, the factor $\sqrt{K}=\sqrt{1.36}$ for the tree-level SM amplitude $A_{\rm SM}$) to fit the $t \bar t$
SM production cross section with LO calculations.
\end{itemize}
For a detailed analysis of the various uncertainties see Ref.~\cite{Frederix:2007gi}.
Nonetheless, most of our results for
$d \sigma/dM_{t {\bar t}}$ are at least marginally compatible with the CDF results,
within $2 \sigma$.


In order to make a more detailed inspection, in Fig.~\ref{fig:AFBLH} we present the
$M_{t {\bar t}}$-restricted forward-backward asymmetries, $i.e.$, those
calculated by the expression (\ref{Afb}) where the phase space in the
numerator and the denominator is restricted by
$M_{t {\bar t}} < M_{t {\bar t}}^{\rm edge}$ (the quantity
$A_{\rm FB}^{t,{\rm low}}$)
or by $M_{t {\bar t}} > M_{t {\bar t}}^{\rm edge}$
(the quantity $A_{\rm FB}^{t,{\rm high}}$).
These asymmetries were calculated for the two aforementioned choices of
parameter values $(\lambda,d_{\U})$, and are compared in the Figure
with the CDF measured values \cite{CDF9724} subtracted by the
(LO+NLO+NLL) SM values \cite{Almeida:2008ug}. This subtraction is needed for comparison with
our results, for the same reason as in the (unrestricted) $A^t_{\rm FB}$
of Eq.~(\ref{Afbour}),
\begin{equation}
A_{\rm FB}^{t,{\rm X}} = A_{\rm FB}^{\rm X}({\rm exp}) - A_{\rm FB}^{\rm X}({\rm SM}) \  \qquad ({\rm X}
= {\rm low}, {\rm high}) \ .
\label{AtX}
\end{equation}
We see that the experimental uncertainties are very
large, especially for $A_{\rm FB}^{t,{\rm high}}$ at
$M_{t {\bar t}}^{\rm edge}=600$ GeV or higher. Nonetheless, the central
experimental values appear to suggest the fall of $A_{\rm FB}^{t,{\rm high}}$
as a function of  $M_{t {\bar t}}^{\rm edge}$ at
$M_{t {\bar t}}^{\rm edge} > 600$ GeV.
It is interesting that a model involving axigluon
exchange does give such a fall for at least one (benchmark) choice of
parameters (with: $g^q_A = -g^t_A$) \cite{Frampton:2009rk}.
On the other hand, our model does not show such a behavior. This issue
remains inconclusive because of: (i) the aforementioned very large
experimental uncertainties of $A_{\rm FB}^{\rm high}$ at high $M_{t {\bar t}}^{\rm edge}$; (ii) the severely restricted phase space at high
$M_{t {\bar t}}^{\rm edge}$. Namely, our simplified approach of rescaling
the tree-level SM (QCD) amplitude by a fixed factor
($\sqrt{K}=\sqrt{1.36}$) for all $M_{t {\bar t}}$ values becomes
increasingly unreliable when $M_{t {\bar t}}^{\rm edge}$ increases in
$A_{\rm FB}^{t,{\rm high}}$, because the phase space becomes so severely restricted.

In conclusion, we investigated whether colored flavor-conserving unparticle
physics can explain the measured forward-backward asymmetry value
for the $t {\bar t}$ production at the Tevatron; the latter measured value shows
$2 \sigma$ deviation from the QCD SM value.
{\it With a natural assumption of quark flavor-blind and chirality-independent
interactions to unparticle},  our calculations
indicate that
the aforementioned unparticle contributions can explain this deviation.
We found an area of the (two-)parameter space of the unparticle physics which
gives the results compatible with the measurements of the
forward-backward asymmetry and of the total cross section for the
$t {\bar t}$ production at the Tevatron. The resulting values of
the differential $d\sigma/dM_{t\bar t}$ cross section and the
$M_{t {\bar t}}$-restricted forward-backward asymmetries
are only marginally compatible with the measured values.


\begin{acknowledgments}
\noindent
C.H.C was supported in part by the National Science Council of R.O.C. under Grant No. NSC-97-2112-M-006-001-MY3.
G.C. was supported in part by FONDECYT (Chile) Grant No.~1095196 and Rings Project (Chile) ACT119.
The work of C.S.K.  was supported in part by the Basic Science
Research Program through the NRF of Korea funded by MOEST
(2009-0088395), in part by KOSEF through the Joint Research
Program (F01-2009-000-10031-0).
\end{acknowledgments}


\end{document}